\documentclass{article}
\usepackage{calligra}
\usepackage{epsfig}
\usepackage{epstopdf}
\usepackage{graphicx}
\usepackage{amsmath}
\usepackage{amsfonts}
\usepackage{dsfont}
\usepackage{xcolor}
\usepackage{slashed}
\usepackage{amssymb}
\usepackage{bbold}
\usepackage{cprotect}
\usepackage{bm}
\usepackage{mathtools}
\usepackage[mathscr]{euscript}
\usepackage{jcappub}

\def\rd{{\rm d}}

\def\erf{{\rm erf}}
\def\bB{{\bf B}}
\def\bk{{\bf k}}
\def\br{{\bf r}}

\newcommand{\red}[1]{\textcolor{black}{#1}}

\title{On the contribution of galaxies \red{to the magnetic field in cosmic voids}}
\author{K\'aroly Seller,}
\emailAdd{karoly.seller@uni-hamburg.de}
\author{G\"unter Sigl}
\emailAdd{guenter.sigl@desy.de}
\affiliation{II. Institut f\"ur Theoretische Physik, University of Hamburg,\\
Luruper Chaussee 149, 22761 Hamburg}

\abstract{Astrophysical processes can contribute to magnetic fields within cosmic voids
either through magnetized outflows from the astrophysical large-scale structure or through
superposition of dipolar contributions from individual galaxies. Such astrophysical
magnetic fields represent a foreground to possible space-filling primordial magnetic
fields seeded in the early Universe. In this paper, we provide a qualitative description of the 
screening of magnetic fields by intergalactic plasmas. We find that contributions from
superposition of static dipoles are highly suppressed and cannot explain indications
for lower bounds based on observations of $\gamma$--ray cascades from high energy
sources such as blazars.}

\begin{document}
\maketitle

\noindent

\section{Introduction}
The large-scale distribution of cosmic magnetic fields is neither experimentally mapped out in large detail, nor completely understood theoretically~\cite{Beck:2015jta,Shukurov_Subramanian_2021,Durrer:2013pga,Subramanian:2015lua,Vachaspati:2020blt}. 
This is \red{particularly} true for magnetic fields in cosmic voids, which are thought to be rather pristine and uncontaminated by astrophysical ``pollution''. 
\red{Such voids could therefore serve as potential probes of primordial magnetic fields created in the early Universe}, see some of the reviews cited above.

Space-filling, large-scale cosmic magnetic fields can be bound from above through their effect on the cosmic microwave background, which limits them to below about 1 nG~\cite{Minoda:2020bod}. 
Furthermore, the deflection of \red{ultra-high-energy} cosmic rays from the direction of the Perseus-Pisces galaxy cluster \red{has} been used to constrain the magnetic field strength \red{in the void between this source and Earth to below} $\sim10^{-10}\,$G~\cite{Neronov:2021xua}.

%On the other hand, electromagnetic cascades at $\mathcal{O}(10-100)$ GeV from active galactic nuclei (AGN) sources of $\gamma$-rays are influenced by magnetic fields in cosmic voids, either through (i) time delays of bursts~\cite{Plaga:1995ins}, (ii) an angular broadening of the source image~\cite{Elyiv:2009bx}, or (iii) through the spectrum of the $\gamma$-ray sources~\cite{Neronov:2010gir}. 
\red{Electromagnetic cascades at $\mathcal{O}(10-100)$ GeV, initiated by $\gamma$--rays from active galactic nuclei, are influenced by magnetic fields in cosmic voids, leading to (i) time delays of bursts~\cite{Plaga:1995ins}, (ii) angular broadening of the source images~\cite{Elyiv:2009bx}, and (iii) change in the observed photon spectrum~\cite{Neronov:2010gir}.}
Based on these, over the last 15 years \red{multiple} lower limits and estimates of magnetic field strengths in voids have been claimed~\cite{Neronov:2010gir,Taylor:2011bn,Dolag:2010ni,Essey:2010nd,Finke:2015ona,Fermi-LAT:2018jdy,AlvesBatista:2020oio,MAGIC:2022piy,HESS:2023zwb,Vovk:2023qfk,Tjemsland:2023hmj,Blunier:2025ddu,Webar:2025qbp}.
Under conservative assumptions on the duty cycle of the sources the currently strongest of those lower bounds can be approximated as~\cite{MAGIC:2022piy}
\begin{equation}
    B \gtrsim \left\{
    \begin{array}{lr}
        1.8 \times 10^{-17}~\mathrm{G}\,, & \lambda_B > 0.2~\mathrm{Mpc} \\
        1.8 \times 10^{-17} \left( \lambda_B / 0.2~\mathrm{Mpc} \right)^{-1/2}~\mathrm{G}\,, & \lambda_B < 0.2~\mathrm{Mpc} \\
    \end{array}
    \right.\,.
    \label{eq::igmf_limit}
\end{equation}

We note in passing that \red{the lower bounds in \eqref{eq::igmf_limit}} may be modified by possible plasma instabilities acting on the electron--positron pairs in the electromagnetic cascades \cite{Broderick:2011av}. 
Such instabilities could provide at least a partial alternative to magnetic fields \red{in} explaining the $\gamma$--ray observations. 
As recent investigations suggest that such effects are small \red{(}see, e.g., Ref.~\cite{Alawashra:2024ndp}\red{)}, we will ignore them here, although the case is not completely settled yet~\cite{Alawashra:2025tcj}.

It is an interesting question whether the magnetic fields suggested by the lower bounds mentioned above can be explained by purely astrophysical processes\red{.}
\red{One possible explanation is that} magnetic fields \red{are} transported \red{into the voids through galaxy and quasar outflows}~\cite{Furlanetto:2001gx,Garcia:2020kxm,Aramburo-Garcia:2022gzn}. 
It has been found in these studies that the \red{volume} filling factor of fields of \red{sufficient} strength is unlikely to \red{exceed} $\sim20\%$, \red{unless a homogeneous (and thus primordial) magnetic field of comparable strength was already present initially}.

\red{An alternative explanation that} has recently been claimed \red{argues} that magnetic fields in the strength range of \eqref{eq::igmf_limit} do not have to be primordial, but could simply be explained by the superposition of magnetic dipoles of individual galaxies~\cite{Garg:2025mcc}. 
However, this \red{requires} a magnetostatic approximation in the absence of plasma. 
The presence of highly ionized plasmas throughout the Universe \red{produces} screening effects that can substantially suppress magnetic fields at large distances from their sources.

Motivated by this, in the present paper we provide a qualitative description of the screening of magnetic fields by intergalactic plasmas.
The results presented here are in part based on Ref.~\cite{Jedamzik:2010cy}.

\section{Screened magnetic field}

In Ref.~\cite{Jedamzik:2010cy} the evolution of the magnetic field is discussed in plasmas where the plasma flow is neglected.
This can be a reasonable first approximation for non-relativistic plasmas that constitute the intergalactic medium.
With this approximation, in Fourier space the evolution equation reads:
\begin{equation}
    \partial_z\bB_\bk(z)=\frac{2\bB_\bk(z)}{1+z}+\eta(z)\frac{1+z}{H(z)}\bk^2\big[\bB_\bk(z)-4\pi{\bf M}_\bk^\perp(z)\big]\,,
\end{equation}
where $z$ is the redshift, $H(z)$ is the Hubble rate and $\eta(z)$ is the resistivity. 
The orthogonal part of the magnetization vector-field is defined as ${\bf M}_\bk^\perp = {\bf M}_\bk-\big(\hat\bk\cdot{\bf M}_\bk\big)\hat\bk\,,$ with $\hat\bk\equiv\bk/|\bk|$.
The analytic solution to this differential equation is \cite{Jedamzik:2010cy}
\begin{equation}
\label{eq:Bk_solution_integral}
    \bB_\bk(z)=4\pi\bk^2(1+z)^2\int_z^\infty\rd z'\,\exp\bigg[-\bk^2\int_z^{z'}\rd\tilde z\,\frac{\eta(\tilde z)(1+\tilde z)}{H(\tilde z)}\bigg]\frac{\eta(z'){\bf M}_\bk^\perp(z')}{(1+z')H(z')}\,.
\end{equation}
With proper assumptions this nested integral is easily solvable without much loss of generality.
The assumptions and approximations we use in this paper are as follows:
\begin{enumerate}
    \item[(a)] We assume that the resistivity only depends on the temperature through a simple power law, $\eta(T)\propto T^{-n_{\eta}}$. 
    We may rewrite this to redshift by using the relation $T_{\rm p}(z)=T_{\rm p,0}(1+z)^{n_T}$, where the value of $n_T$ depends on the nature of the plasma (i.e., relativistic or non-relativistic).
    The general evolution of the resistivity is then
    \begin{equation}
        \eta(z)=\eta_0(1+z)^{-n_\eta n_T}\,,
    \end{equation}
    where $\eta_0$ is the resistivity of the plasma today.
    As a specific example, we will use the Spitzer resistivity (see, e.g., \cite{Choudhuri_1998}), where $n_\eta=3/2$ and
    \begin{equation}
        \label{eq:Spitzer_resistivity}
        \eta_{\rm Sp}(z)=C\frac{e^2\sqrt{m_e}}{T_e^{3/2}} = C\frac{e^2\sqrt{m_e}}{T_{e,0}^{3/2}}(1+z)^{-3n_T/2}\,.
    \end{equation}
    An undetermined dimensionless constant $C\sim10$ is used as a fudge factor to account for uncertainties in plasma screening strengths.
    Moreover, in natural units $e^2\approx4\pi/137$, $m_e\approx 511\,{\rm keV}$ is the electron mass, and $T_{e,0}$ is the temperature of the electron-plasma today.. 
    \item[(b)] We assume that the Universe is in a single-component dominated era (e.g., dark energy- or matter-domination).
    The Hubble parameter is then defined as
    \begin{equation}
        H(z)=H_0(1+z)^{n_H}\,,
    \end{equation}
    where $H_0\approx 1.4\times 10^{-33}\,{\rm eV}$ is the Hubble parameter today. 
    Analytic results only exist in this approximation, however, using a realistic Hubble parameter does not effect the qualitative results presented later.
    \item[(c)] We approximate the $z$-scaling of the magnetization as
    \begin{equation}
        {\bf M}_\bk^\perp(z)\simeq(1+z)^2{\bf M}_\bk^\perp(0)\,.
    \end{equation}
    This follows from the magnetic energy density scaling as $E_{\rm M}\propto |\bB_\bk|^2 \propto (1+z)^{-4}$\,.
\end{enumerate}
The values of the exponents $n_\eta$, $n_T$ and $n_H$ depend on the properties of the intergalactic medium and the era of the Universe.
In particular, one has $n_T=1~(n_T=2)$ for relativistic (non-relativistic) plasmas, while $n_H=3/2~(n_H=0)$ for a matter (cosmological constant) dominated Universe.
The radiation dominated Universe is not a relevant era for structure formation thus we do not consider it.
Regardless, the actual values of $\{n_\eta,n_T,n_H\}$ are largely irrelevant for the qualitative behavior of $\bB_\bk$.

A useful quantity, introduced in Ref.~\cite{Jedamzik:2010cy}, is the magnetic screening length $l_{r,0}=2\pi/k_{r}(0)$, where the critical wave-number $k_r$ is defined via
\begin{equation}
    k_r(z)=\bigg[\int_z^{\infty}\rd\tilde z\,\frac{\eta(\tilde z)(1+\tilde z)}{H(\tilde z)}\bigg]^{-1/2}\,.
\end{equation}
Notice that in \eqref{eq:Bk_solution_integral} the inner integrand gives roughly $\sim\exp(-k^2/k_r^2)$, which provides a suppression for high-$k$ modes that is characteristic of screening.
Using our prior assumptions (a) and (b) we find in general that the critical wave-number is
\begin{equation}
    \label{eq:kr_expression}
    k_r(z)=\sqrt{\frac{H_0}{\eta_0}}\Big[\frac{(1+z)^{2-n_H-n_\eta n_T}}{n_H+n_\eta n_T-2}\Big]^{-1/2}\,,\qquad{\rm if}~n_H+n_\eta n_T>2\,.
\end{equation}
Integrating to the present day ($z=0$), up to $\mathcal{O}(1)$ factors one has $k_{r}(0)\sim \sqrt{H_0/\eta_0}$.
As an example, let us look at the Spitzer resistivity given in \eqref{eq:Spitzer_resistivity}, in a non-relativistic plasma ($n_T=2$) at $T_{e,0}=10~{\rm eV}$ in a matter-dominated Universe ($n_H=3/2$). 
Today, at $z=0$ the screening scale is
\begin{equation}
    k_{r,0}\equiv k_r(0)\approx 10^{-17}~{\rm eV}\approx 3.6\times10^{4}~{\rm pc}^{-1}\,.
\end{equation}
The screening length is extremely small compared to the astrophysical distances relevant in voids,
\begin{equation}
    \label{eq:Spitzer_lr0}
    l_{r,0}=\frac{2\pi}{k_{r,0}}\approx 2\times 10^{-4}~{\rm pc}\ll \mathcal{O}(10^{6})~{\rm pc}\sim l_{\rm void}\,.
\end{equation}
Consequently, we do not expect that sufficiently strong magnetic fields are able to propagate to $\mathcal{O}({\rm Mpc})$ distances within voids to reproduce lower bounds given in \eqref{eq::igmf_limit}.

To make our argument complete, we return to \eqref{eq:Bk_solution_integral} and integrate it analytically using all three assumptions (a)--(c) to find explicit formulae for the field strength.
At $z=0$ we find
\begin{equation}
\label{eq:Bk_solution_analytic}
    \bB_\bk(0)=4\pi\bigg[1-\exp\bigg(-\frac{k^2}{k_{r,0}^2}\bigg)\bigg]{\bf M}_\bk^\perp(0)\,.
\end{equation}
This formula is the same irrespective of what $n_H$, $n_T$, and $n_\eta$ are, at least as long as the integrals remain finite at the chosen values (see the condition in \eqref{eq:kr_expression} in particular).
These parameter values are only important in the determination of the screening scale $k_{r,0}$, and the presented form for the screened magnetic field in \eqref{eq:Bk_solution_analytic} is in practice quite general.
For small distances $k\gg k_{r,0}$ the exponential term is negligible and we re-obtain the unscreened magnetic field:
\begin{equation}
    \lim_{k\gg k_{r,0}}\bB_\bk\simeq 4\pi{\bf M}_\bk^\perp\,.
\end{equation}
For large distances $k\ll k_{r,0}$ we expect the screening to have considerable effects, and we find
\begin{equation}
    \label{eq:Bk_small_k}
    \lim_{k\ll k_{r,0}}\bB_\bk\simeq 4\pi\frac{k^2}{k_{r,0}^2}{\bf M}_\bk^\perp\,.
\end{equation}
For the Fourier component with wavelength $\lambda\simeq l_{\rm void}\sim1~{\rm Mpc}$ the screening results in a suppression of about $10^{-20}$ compared to the vacuum result if we consider $l_{r,0}$ as in \eqref{eq:Spitzer_lr0}.

To reconnect with Ref.~\cite{Garg:2025mcc}, we now calculate the screened magnetic field of a single magnetic dipole in direct space, used for modeling the magnetic field of a single galaxy.
After a simple Fourier transformation of \eqref{eq:Bk_solution_analytic} the magnetic field of the screened dipole is obtained (in Gaussian units):
\begin{equation}
    \label{eq:B_screened_dipole}
    \bB_{\rm screened-dipole}({\bf u})=-k_{r,0}^3 \frac{{\bf m}u^2\mathscr{F}_{\rm s}(u;2)-3{\bf u}({\bf m}\cdot{\bf u})\mathscr{F}_{\rm s}(u;6)}{u^5}\,,
\end{equation}
where ${\bf u}=k_{r,0}\br$ is the dimensionless distance scale relative to the screening length and we defined the auxiliary function describing screening:
\begin{equation}
    \mathscr{F}_{\rm s}(u;n)=1-\erf\Big(\frac{u}{2}\Big)+\frac{u}{\sqrt{\pi}}{\rm e}^{-u^2/4}\Big(1+\frac{u^2}{n}\Big)\,.
\end{equation}
Here $\erf(x)$ is the (Gaussian) error function.
Note, that the different components (longitudinal and transverse) of the magnetic field are screened slightly differently.

The result in \eqref{eq:B_screened_dipole} clearly resembles the usual vacuum magnetic field of the dipole.
In fact, it exactly reproduces the dipole field in the limit $u\to 0$ (i.e., $r\ll l_{r,0}$):
\begin{equation}
\begin{aligned}
    \lim_{u\to0}\mathscr{F}_{\rm s}(u;n)=1\quad\longrightarrow\quad \lim_{r\ll l_{r,0}}\bB_{\rm screened-dipole}(\br)=\frac{3\br({\bf m}\cdot\br)-{\bf m}r^2}{r^5}\equiv\bB_{\rm dipole}(\br)\,.
\end{aligned}
\end{equation}
However, for large distances relative to the screening length the field decays exponentially fast,
\begin{equation}
    \lim_{u\gg1}\mathscr{F}_{\rm s}(u;n)\simeq\frac{u^3}{n\sqrt{\pi}}{\rm e}^{-u^2/4}~\longrightarrow~\lim_{r\gg l_{r,0}}\bB_{\rm screened-dipole}(\br)\sim\Big(\frac{r}{l_{r,0}}\Big)^3{\rm e}^{-r^2/4l_{r,0}^2}~\bB_{\rm dipole}(\br)\,.
\end{equation}
In summary, the magnetic field is exponentially suppressed for large distances relative to the screening length, while it retains its dipole form for small distances. 

Following Ref.~\cite{Garg:2025mcc}, we estimate the magnetic field strength in voids from the collective effect of numerous galaxies, each modeled as dipoles.
We can safely assume that the size of the galaxies is much smaller than the size of the void and consequently we may approximate the magnetic field of a single galaxy as
\begin{equation}
    \label{eq:void_B1_field_approx}
    B_{\rm galaxy}(r_{\rm void}) \simeq B_0 \,\rm{e}^{-k_{r,0}^2r_{\rm void}^2/4}\,,
\end{equation}
where we use $r_{\rm void}\sim 20\,{\rm Mpc}$ and $B_0\sim 10^{-6}\,{\rm G}$ \cite{Unger:2023lob}.
In general, one expects $l_{r,0} \ll r_{\rm void}$ (see \eqref{eq:Spitzer_lr0} for the Spitzer resistivity with reasonable plasma parameters) and thus an exponentially small $B_{\rm galaxy}(r_{\rm void})$. 
Approximating the number of galaxies around the void by $N_{\rm g}\simeq 4\pi r_{\rm void}^2/d_{\rm ig}^2$, where $d_{\rm ig}\sim 1\,{\rm Mpc}$ is the typical distance between galaxies, the total magnetic field in the middle of the spherical void becomes
\begin{equation}
    B_{\rm void} \simeq B_{\rm galaxy}(r_{\rm void})\sqrt{N_g} \approx (10^{-5}\,{\rm G})\times {\rm e}^{-r_{\rm void}^2/4l_{r,0}^2}\,.
\end{equation}
In order to obtain $B_{\rm void}\sim \mathcal{O}(10^{-16})\,{\rm G}$ one requires a screening length of order $l_{r,0}^{(\rm ideal)}\approx 2\,$Mpc, which is far from realistic for intergalactic plasmas.

Finally, we also mention that in general there is no relationship between the Fourier-component at a specific wave-number $k_0=2\pi/r_0$, and the corresponding magnetic field strength at $r_0$.
Thus, the exponentially small field in \eqref{eq:void_B1_field_approx} is not in any contradiction with the quadratically small field in \eqref{eq:Bk_small_k}.
Regardless, we can safely say that both give magnetic fields that are many-many orders of magnitude too weak to explain the observed lower bounds of \eqref{eq::igmf_limit}.

\section{Conclusions}

In this paper, we have shown that dipole-like galactic magnetic fields cannot account for the observational lower bounds on the magnetic field strength in voids. 
Such fields cannot propagate to cosmic distances because of the presence of intergalactic plasma.
In fact, for reasonable values of plasma resistivity, the resulting field strength falls short of the observed lower bound by many orders of magnitude.
For the Fourier spectrum of the magnetic field ($\bB_\bk$) we find a $(k/k_{r,0})^2$ suppression at small--$k$ modes.
In real space, the screening effect is exponentially strong, effectively eliminating the field on scales $r \gg l_{r,0}$. 
We conclude that a screening length of $l_{r,0} \sim {\rm few~Mpc}$ would be necessary to sustain the simple galactic field explanation. 
However, such values are astrophysically implausible.
Finally, we emphasize that our calculation is a simplified argument showing that galactic fields alone cannot explain the observed void fields.
Other mechanisms discussed in the introduction (e.g., instabilities and outflows) may play an important role and further complicate the picture.

\acknowledgments
We acknowledge support by the Deutsche Forschungsgemeinschaft (DFG, German Research Foundation) under Germany’s Excellence Strategy -- EXC 2121 ``Quantum Universe'' -- 390833306.
We thank Kandaswamy Subramanian, Ruth Durrer, Jennifer Schober, and Deepen Garg for discussions regarding this topic.

\bibliographystyle{JHEP}
{\sloppy\footnotesize
\bibliography{paper}
}

\end{document}